\documentclass[
 preprint,
 longbibliography,
 amsmath,amssymb,
 aps,
 pra,
floatfix,
]{revtex4-1}

\usepackage{graphicx}
\usepackage{dcolumn}
\usepackage{bm}
\usepackage{mathtools}
\usepackage{hyperref}
\usepackage[mathlines]{lineno}
\usepackage{comment}

\begin{document}

\preprint{APS/123-QED}

\title{An active approach to colloidal self-assembly}

\author{S. A. Mallory$^1$, C. Valeriani$^{2}$, A. Cacciuto$^1$}
\email{ac2822@columbia.edu}
\affiliation{$1$Department of Chemistry, Columbia University\\ 3000 Broadway, New York, NY 10027\\ }
\affiliation{$^{2}$Departamento de Fisica Aplicada I, Facultad de Ciencias Fisica, Universidad Complutense de Madrid, 28040 Madrid, Spain}


\date{\today}

\begin{abstract}
In this review we discuss recent advances in the self-assembly of self-propelled  colloidal particles and highlight some of the most exciting results in this field with a specific focus on dry active matter.  We explore this phenomenology through the lens of the complexity of the colloidal building blocks. We begin by considering the behavior of isotropic spherical particles. We then discuss the case of amphiphilic and dipolar Janus particles. Finally, we show how  the geometry of the colloids and/or the directionality of their interactions can be used to control the physical properties of the assembled active aggregates, and suggest possible strategies on how to exploit activity as a tunable driving force for self-assembly.  The unique properties of active colloids lend promise for the design of the next generation of functional, environment-sensing microstructures able to perform specific tasks in an autonomous and targeted manner. 
\end{abstract}

\maketitle

\tableofcontents

\section{\label{sec:intro}Introduction}

The way nature generates functional structures at the microscale is, in principle, extremely simple. This autonomous construction process is called self-assembly, and can be qualitatively understood with the following analogy.  Imagine placing a number of different LEGO bricks into a cardboard box. Now shake that box for a sufficiently long time. Finally, open the box to find a perfectly assembled structure. This seemingly outrageous idea works at the micro and nano scale because, unlike the LEGO bricks which lay motionless unless an external force is applied (i.e. shaking), microscopic components freely and aimlessly diffuse via thermal motion across the space made available to them. Given enough time, they will eventually find each other. The process of self-assembly leverages this thermal motion to generate microstructures at practically no external cost. As such, this bottom-up approach to materials engineering is considered to be a major competitor to other more expensive top-down fabrication methods. 

Unfortunately, the drawback of having to rely on random motion, as opposed to deliberate motion, is that microcomponents will indiscriminately find each other without necessarily matching to the correct partners to form the desired microstructure. This typically leads to a malformed final structure.  For self-assembly to be successful, a very intricate game of labeling must be played. Each component must be encoded with its own local blueprint of where it will reside in the final structure. Clearly, the complexity of the rules (i.e. how the LEGO bricks should stick and fit together) increases with the complexity of the desired final structure. As such, self-assembly is a rather delicate process, and the formation of defect-free structures is hardly achievable unless a careful design of the building blocks is performed beforehand. 

In this short review, we discuss recent advances in the field of Active Matter with a specific focus on the self-assembly of active colloids. Throughout the text, we discuss ways in which self-propulsion can be exploited to form robust macroscopic and mesoscopic structures whose properties can be tuned by the degree of activity. Prior to diving into this topic, it is instructive to provide a brief overview of the field of Active Matter. 

Over the past decade, Active Matter has transformed and energized the fields of Statistical Mechanics and Soft Condensed Matter.  All active systems share the hallmark feature of being composed of self-driven units capable of converting stored or ambient energy into systematic movement. Examples of active systems permeate both the macroscopic and microscopic world and include: human crowds, flocks and herds of animals, living cells and tissues, active colloids and synthetic microswimmers. Active Matter represents a fundamentally new non-equilibrium regime within Statistical Mechanics. In contrast to traditional non-equilibrium systems, where directional driving forces emerge as a result of global changes of the thermodynamic variables or boundary conditions (such as temperature and pressure), active systems are intrinsically out of equilibrium at the single particle level. The combination of these unique non-equilibrium driving forces and the inherently stochastic nature of these systems have endowed active systems with remarkable collective behavior which ranges from bacterial turbulence \cite{dunkel_fluid_2013,wolgemuth_collective_2008} to self-regulation \cite{gopinath_dynamical_2012} and, more generally, leads to large spatial correlations that are typically only observed near critical points in equilibrium systems. For a thorough discussion of these properties, we direct the reader to a number of comprehensive reviews recently written on the subject (see for instance \cite{bechinger_active_2016,wang_one_2015,toner_hydrodynamics_2005,cates_motility-induced_2015,ramaswamy_mechanics_2010,klapp_collective_2016,marchetti_minimal_2016,bialke_active_2015,zottl_emergent_2016,dey_chemically_2017,wang_small_2013,menzel_tuned_2015,marchetti_hydrodynamics_2013,j.ebbens_pursuit_2010,chate_modeling_2008,romanczuk_active_2012,j.ebbens_pursuit_2010} and the references therein).
 
The current rise in popularity of colloidal active matter is due in large part to the recent advances in colloidal synthesis. Through the pioneering work of synthetic chemists and material scientists, there is now a number of experimental realizations  of synthetic microswimmers and active colloids. These active particles can be thought of as synthetic analogs of swimming bacteria.  However, a major benefit of these synthetic variants is that, unlike bacteria, it is possible to systematically tailor inter-particle interactions and modulate their swimming velocity. For details about the growing library of synthetic microswimmers and their associated propulsion mechanisms, we direct the readers to several recent reviews on the subject \cite{wang_one_2015,wang_small_2013,bechinger_active_2016,j.ebbens_pursuit_2010}.  

The functionality of these active colloids and synthetic microswimmers makes them the ideal tool to manipulate matter at the microscale. A synthetic microswimmer's ability to autonomously navigate complex microfluidic environments conjures up a host of appealing applications, which include targeted drug delivery to specific cells \cite{din_synchronized_2016},  information storage and computation \cite{woodhouse_active_2017},  clean-up and neutralization of environmental pollutants \cite{ebbens_active_2016}, self-propelled nanotools \cite{solovev_self-propelled_2012},  and the massive parallel assembly of microscopic structures \cite{mallory_activity-assisted_2016}. These potential applications are built around the unique self-driven nature of the microswimmers and their ability to manipulate, sense, and transport material at the microscale. 

Broadly speaking, it is fair to say that the field of self-assembly is still in its infancy. Many aspects of the process are still not well understood, and we have yet to reach the stage where we can make reliable robust predictions for materials engineering.\cite{sacanna_shaping_2013,glotzer_self-assembly:_2004,jiang_janus_2010,klapp_collective_2016,ravaine_synthesis_nodate,sacanna_shape-anisotropic_2011,yi_recent_2013,chen_janus_2012,pawar_fabrication_2010,sacanna_lock_2010,whitelam_role_2009}. However, there is now clear evidence that the formation of defect-free structures in non-active systems is most probable when the rate of association of the particles from the bulk into the target aggregate is comparable to that of their dissociation from it, making the process tediously slow.  Given the circumstances, the question of whether self-propulsion can be used as an extra handle to control and speed up the self-assembly of colloidal particles becomes extremely relevant. There are two open research areas within active self-assembly: finding ways to impart dynamic functionality to self-assembled structures and using activity to improve self-assembly. Here, we attempt to extract from the vast amount of phenomenology that has recently been reported, key features that may be important for improving self-assembly.

\section{\label{sec:active-colloids} Active Colloids}

At the microscale in the absence of external forces, the motion of a passive colloid is driven by equilibrium thermal fluctuations originating from the solvent. The dynamics can be described by the theory of Brownian motion, and much of the collective behavior can be explained within the framework of equilibrium Statistical Mechanics.  In contrast, active colloids autonomously convert energy available in the environment, whether it be chemical \cite{howse_self-motile_2007}, electromagnetic \cite{bricard_emergence_2013}, acoustic \cite{ahmed_self-assembly_2014}, or thermal energy \cite{jiang_janus_2010}, into directed mechanical motion. In this case, the swimming direction is not determined by an external force at a global level, but is an intrinsic local property of the individual swimmers.  Thus, it is often noted that active fluids are driven out of equilibrium at the single particle level.  

Active colloids, and more generally all colloidal particles, move in a low Reynold's number environment, (i.e. viscous forces from the surrounding fluid dominate over any inertia forces) \cite{lauga_hydrodynamics_2009}, making it very challenging to develop propulsion mechanisms for active colloids. Since inertia  does not contribute to the motion of the colloid, a constant force must be applied to the colloid to maintain its swimming velocity. A common scheme to achieve this is  creating and sustaining some sort of gradient across the colloid.  This gradient driven motion is known as phoretic transport and the current belief is that if a particle can generate its own local field or gradient (whether it be chemical, electromagnetic, or thermal), then the particle will be propelled forward via "self-phoresis" \cite{j.ebbens_pursuit_2010,howse_self-motile_2007,golestanian_propulsion_2005,golestanian_designing_2007}.  

\begin{figure}[h]
 \centering
    \includegraphics[width=0.9\textwidth]{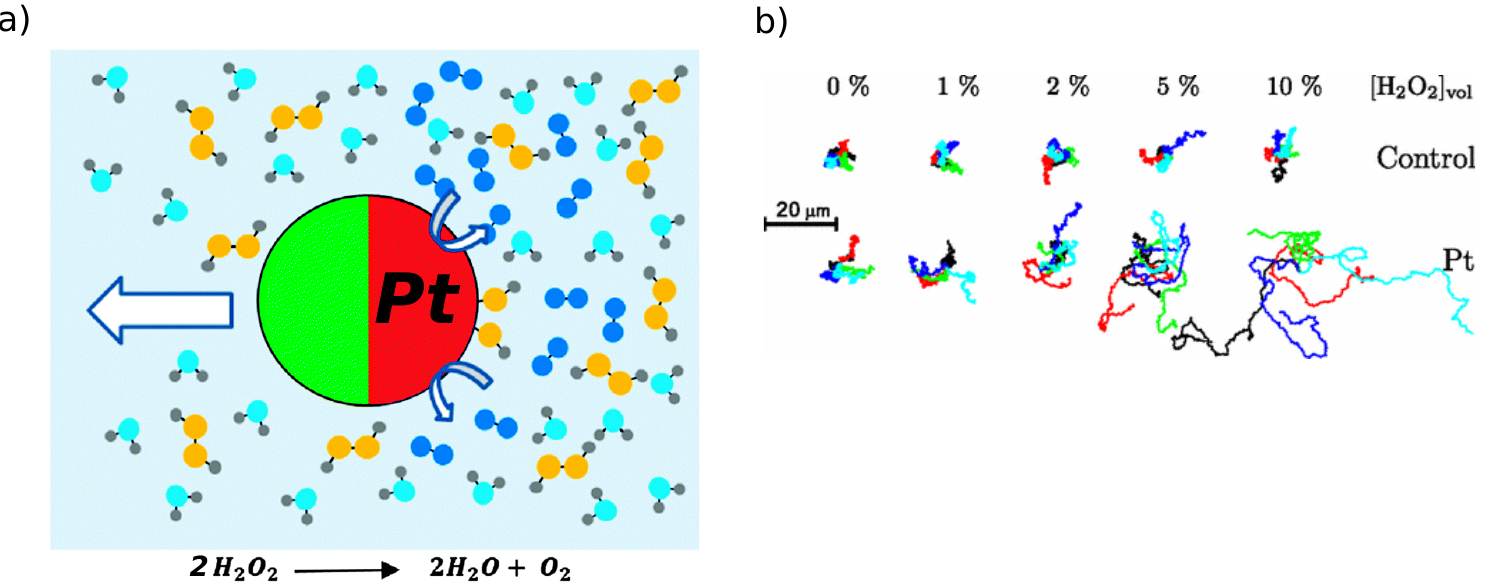}
  \caption{(a) Sketch of the propulsion mechanism for the active colloid synthesized by Howse et al. \cite{howse_self-motile_2007}.  The platinum cap (shown in red) decomposes hydrogen peroxide into water (light blue) and oxygen (dark blue) resulting in an oxygen gradient across the particle. Panel \textit{a} was adapted from ref \cite{j.archer_glancing_2015} with permission from The Royal Society of Chemistry. (b) Typical trajectories of the active colloid synthesized by Howse et al. at various hydrogen peroxide fuel concentrations.  Panel \textit{b} reproduced with permission from American Physical Society.}
  \label{fig:fig1}
\end{figure}

Here, we briefly introduce one of the simplest propulsion mechanisms for active colloids.  This particular type of active colloid is Janus in nature, i.e. the colloid itself is composed of two different materials.  As shown in Fig.\ref{fig:fig1}a, the active colloid synthesized by Howse et al. \cite{howse_self-motile_2007} consists of a spherical polystyrene bead with a thin layer of platinum (Pt) deposited on one hemisphere.  Platinum was chosen for the catalytic cap as it readily decomposes the aqueous hydrogen peroxide fuel source into water and oxygen. As the hydrogen peroxide is decomposed by the catalytic Pt cap, a concentration gradient of oxygen is generated and sustained across the colloid resulting in its self-propulsion. Although the exact details of the propulsion mechanism even in this simple case are not fully understood, and competing mechanisms of propulsion have been suggested\cite{brown_ionic_2014}, there is consensus that sustaining a gradient across the particle is required to propel these building blocks. For additional details on a variety of self-propelling mechanisms, we direct the reader to several comprehensive reviews in this field \cite{wang_one_2015,wang_small_2013,bechinger_active_2016,j.ebbens_pursuit_2010,zottl_emergent_2016,ebbens_active_2016}.

\subsection{\label{sec:apb_model} Active Brownian Particle Model }

Much of the existing work on Active Matter has relied on simplified or minimal models. One of the most heavily studied is the Vicsek model introduced by Tamas Vicsek \cite{vicsek_novel_1995}. The Vicsek model garnered considerable interest because of its applicability to a wide range of biological systems involving flocking, clustering and migration, see for instance references \cite{chate_modeling_2008,toner_hydrodynamics_2005,ginelli_large-scale_2010,chate_collective_2008}.  With the recent proliferation of synthetic microswimmers, a new model, often referred to as the Active Brownian Particle (ABP) model, \cite{cates_when_2013,khatami_active_2016,solon_active_2015,stenhammar_phase_2014,speck_collective_2016,stenhammar_continuum_2013,bechinger_active_2016,richard_nucleation_2016,toner_following_2016,bialke_negative_2015}, has recently achieved great popularity and will be the main focus of this review.  In the context of the ABP model, active colloids are described as Brownian particles under the additional influence of a constant self-propelling force applied along a fixed particle axis. While this simple representation captures many features of the basic motion of active particles, it neglects often crucial hydrodynamic interactions \cite{matas-navarro_hydrodynamic_2014,matasnavarro_clustering_2015, zottl_hydrodynamics_2014,lauga_hydrodynamics_2009,alarcon_morphology_2017,delfau_collective_2016,theers_modeling_2016,thutupalli_swarming_2011,elgeti_physics_2015,gompper_multi-particle_2009}. Using previously introduced conventions \cite{marchetti_hydrodynamics_2013,steffenoni_microscopic_2017}, the ABP model is a “dry” model of Active Matter, in contrast to “wet” models  where the momentum exchanged between active particles and the surrounding solvent is considered explicitly. 

A spherical active colloid in the APB model is described as a sphere of mass $m$ and diameter $\sigma$ that undergoes Brownian dynamics at a constant temperature $T$, and is subject to an axial propelling velocity of magnitude $v_p$. The dynamics of the ABP evolves following the coupled overdamped Langevin equations for the translational and rotational motion, respectively:
\begin{equation}
\label{tra}
\dot{\pmb{r}}(t) = \frac{1}{\gamma} \pmb{F}(\{r_{ij}\}) + v_p \, \pmb{n} + \sqrt{2D}\,\pmb{\xi}(t)
\end{equation}

\begin{equation}
\label{rot}
\dot{\pmb{n}}(t) = \frac{1}{\gamma_r}\pmb{T}(\{r_{ij}\}) + \sqrt{2D_r}\, \pmb{\xi}(t) \times \pmb{n}
\end{equation}

\noindent Here $v_p$  is directed along a predefined orientation unit vector $\pmb{n}$ which passes through the origin of each particle and connects its poles. The translational diffusion coefficient $D$ is related to the temperature $T$ and the translational friction $\gamma$ via the Stokes-Einstein relation $D=k_bT/\gamma$. It is typical to assume that the rotational diffusion coefficient $D_r$ satisfies the relation $D_r = (3D)/\sigma^2$, which holds true for spherical particles at low Reynold's numbers. The solvent induced Gaussian white-noise terms for both the translational and rotational motion are characterized by $\langle \xi_i(t)\rangle = 0$ and $\langle \xi_i(t) \xi_j(t^\prime)\rangle = \delta_{ij}\delta(t-t^\prime)$. The conservative inter-particle forces and torques acting on the colloid are indicated as $\pmb{F}(\{r_{ij}\})$ and  $\pmb{T}(\{r_{ij}\})$, respectively, and can be computed as 
\begin{equation}
\pmb{F}(\{r_{ij}\}) = \sum_j \pmb{F}_{ij} = -\sum_j\frac{\partial U_{ij}}{\partial\pmb{r}_{ij}}
\end{equation}
and 
\begin{equation}
\pmb{T}(\{r_{ij}\}) = \sum_j \pmb{T}_{ij} = -\sum_j\pmb{\nabla}_{\pmb{n}_{ij}} U_{ij} 
\end{equation}
where $U_{ij}$ is the interaction potential between the colloids.

A very important length-scale in these systems is the persistence length of the active drive defined as $l_p=\frac{v_p \sigma}{D_r}$. This can be related to the Peclet number ${\rm Pe}=\frac{v_p\sigma}{D}=\frac{1}{3}\frac{l_p}{\sigma}$. For a passive colloid, the motion is completely Brownian and the mean square displacement in two dimensions is given by $\langle \pmb{r}^2(t) \rangle = 4Dt$. For an active colloid, the mean square displacement includes an additional term and is given by $\langle \pmb{r}^2(t) \rangle = 4Dt + 2l_p^2[D_rt+1-e^{-D_rt}]$ \cite{hagen_brownian_2011}. For sufficiently large $v_p$, the motion of the active colloid becomes increasingly ballistic at short time scales. However due to the colloid's stochastic rotation, the motion is always diffusive at longer time scales. Typical trajectories of a single active colloid at different fuel concentrations are given in Fig.\ref{fig:fig1}b. Much of the interesting behavior in active systems arises when $l_p$ is significantly larger than the  colloidal diameter.  For instance, one of the most important features of active particles is their ability to exert a unique force or pressure on their surroundings.  This is an area of research that has received considerable interest and a number of analytical, numerical, and experimental results have been published on the subject \cite{zottl_emergent_2016,solon_pressure_2015-1,lee_active_2013,mallory_anomalous_2014,takatori_swim_2014,solon_pressure_2015,takatori_swim_2014,ginot_nonequilibrium_2015,smallenburg_swim_2015,ni_tunable_2015,speck_ideal_2016,marconi_pressure_2016}. In fact, it has been shown that active fluids not only can generate a pressure on a boundary, but they can also induce exotic (depletion-like) effective forces between passive particles (tracers) with range and strength controlled by $l_p$, that can be orders of magnitude larger than what expected for passive systems \cite{leite_depletion_2016,li_brush_2015,ni_tunable_2015,parra-rojas_casimir_2014,ray_casimir_2014,angelani_effective_2011,kaiser_unusual_2014,harder_activity-induced_2014,harder_role_2014,valeriani_colloids_2011}. These results suggest ways of designing microstructures to perform specific tasks when immersed in an active suspension. Most notably, work on how to drive microscopic gears and motors \cite{sokolov_swimming_2010,di_leonardo_bacterial_2010,angelani_self-starting_2009,maggi_self-assembly_2016}, on how to transport colloidal cargos \cite{koumakis_targeted_2013}, how to capture and rectify the motion of active particles \cite{lambert_collective_2010,pototsky_rectification_2013,wan_rectification_2008}, and   how to use  active suspensions to propel wedgelike carriers \cite{kaiser_mechanisms_2015,kaiser_motion_2015,mallory_curvature-induced_2014,mallory_anomalous_2015,angelani_geometrically_2010} has been carried out. In all these instances, the local geometry of these microdevices is crucial to effectively focus the random active motion of the colloids  into directed mechanical work \cite{lee_active_2013,wensink_controlling_2014,fily_dynamics_2014,fily_dynamics_2015,elgeti_wall_2013}.

\section{\label{sec:self-assembly} Self-assembly with ABPs}

In this section, we provide an overview of the various contributions to the field of active colloidal self-assembly. Special attention will be paid to the different types of colloidal interactions, and how they lead to a variety of ordered and disordered structures.  Most of the discussion will be framed within the context of the Active Brownian Particle model, which was introduced in the previous section. The discussion will mainly be confined to two dimensional or quasi-two dimensional systems, as most experiments are performed under these conditions. The aim of this section is to provide some degree of insight into the interplay between activity and particle interactions, and how this relates to the successful self-assembly of a wide range of passive and active microstructures. 

We begin by reviewing the self-assembly behavior of purely repulsive active colloids. After that, we consider a few cases where active colloids interact via a generic isotropic attraction. We conclude this section by considering non isotropic potentials and by discussing recent work on the self-assembly of dipolar Janus ABPs  and amphiphilic Janus ABPs.

\subsection{Isotropically purely repulsive ABPs}

\begin{figure}[h!]
 \centering
    \includegraphics[width=0.5\textwidth]{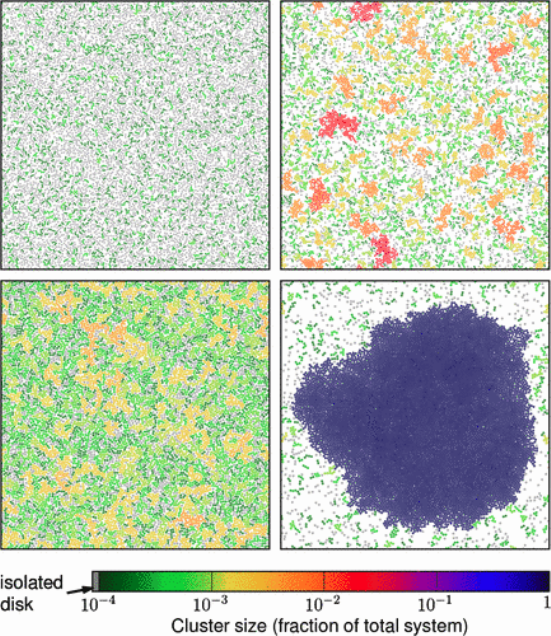}
  \caption{Simulation snapshots of purely repulsive ABPs at two different volume fractions $\phi=0.39$ (top row) and $\phi=0.7$ (bottom row).  The left column are snapshots from a  system of passive particles while the right column shows the corresponding phase in the limit of very large self-propulsions.  Particles are colored according to cluster size. Image reproduced with permission from American Physical Society from ref \cite{fily_athermal_2012}}.
  \label{fig:fig2}
\end{figure}

Self-assembly of passive spherical colloids with short-range isotropic interactions  is typically limited, in terms of structural variety, to closed packed hexagonal crystals. The isotropic nature of the interaction potential does not allow for any close control in either the relative location or orientation of the colloids in a particular aggregate, precluding the formation of more complex structures.  The situation is even more drastic for colloids that interact solely through excluded volume interactions. In fact, at low densities a suspension of purely repulsive spheres exists only in the fluid phase, and when the density becomes sufficiently large  an ordered phase develops as a result of an increased configurational entropy. 

The  behavior becomes much richer whenever some degree of activity is introduced. One of the most interesting and heavily studied phenomena, which appears to be ubiquitous in colloidal active systems, is giant, long-lived density fluctuations and anomalous clustering. A passive suspension of $N$ colloids in a volume $V$ exhibits number fluctuations that scale as $\Delta N\sim \sqrt{N}$ for $N \rightarrow \infty$. Interestingly, in a variety of active systems, $\Delta N$, can become very large and scales as $\Delta N\sim N^a$, where $a$ is an exponent predicted to be as large as one in two dimensions \cite{toner_hydrodynamics_2005,toner_long-range_1995}. This scaling prediction has been verified experimentally \cite{narayan_long-lived_2007,deseigne_collective_2010,peruani_collective_2012} and  in agent-based simulations \cite{chate_simple_2006,chate_modeling_2008}. In these earlier works,  the focus was on elongated active particles (inspired by swimming bacteria and animal flocking) and the associated giant number fluctuations were believed to be induced by the broken orientational symmetry of the particle. More recently, it has been shown that this is a more general phenomenon and has been observed in suspensions of purely repulsive spheres with no alignment interaction \cite{fily_athermal_2012,redner_structure_2013}.  

A related phenomenon observed in suspensions of purely repulsive ABPs is the existence of an athermal motility-induced phase separation (MIPS) \cite{cates_motility-induced_2015,stenhammar_phase_2014,stenhammar_activity-induced_2015,tailleur_statistical_2008,barre_motility-induced_2015,gonnella_motility-induced_2015,speck_effective_2014,buttinoni_dynamical_2013,wysocki_cooperative_2014,marchetti_minimal_2016}.  Here, the system is able to undergo a transition from a dilute fluid to a macroscopic dense phase, which takes place at densities much lower than those expected for the configurational-entropy-driven transition of the passive parent system. Figure \ref{fig:fig2} illustrates this novel phase transition in a simulation of purely repulsive ABPs. The discovery of MIPS  has led to extensive theoretical, experimental, and numerical studies, which have culminated in a number of review articles on the subject \cite{bechinger_active_2016,cates_motility-induced_2015}. This phase transition differs significantly from the ordered states of nematic and polar active systems, which are intimately linked to the broken orientational symmetry of the particle, because spheres lack a mutual alignment mechanism and thus are unable to exchange angular momentum. This phase separation is strikingly similar to the vapor-liquid spinodal decomposition observed in many equilibrium systems, and there is now a significant amount of evidence supporting the hypothesis by Cates and collaborators \cite{aranson_comment_2008,cates_arrested_2010,tailleur_statistical_2008,ball_particle_2013,cates_when_2013} that MIPS is a generic features of systems that are driven out of equilibrium by a persistent local energy input that breaks detailed balance. 
 
The physical mechanism responsible for giant number fluctuations and MIPS can be easily contextualized within a simple kinetic model \cite{redner_structure_2013,fily_athermal_2012} which focuses on the local dynamics at the interface of a forming cluster. At any given time, the size of a cluster  is determined by the flux balance of incoming and outgoing particles.  The time required for an active colloid to leave the surface of a cluster is independent of the propelling velocity $v_p$ and is completely determined by its average rotational diffusion time $t_r \sim 1/D_r$, the collision rate between free colloids and the forming cluster increases linearly with the particle density and propelling speed $v_p$\cite{redner_reentrant_2013}.

By modulating the value of $D_r$, it is possible to control the size of the number fluctuations and retard or promote particles' aggregation \cite{prymidis_state_2016}.   This is an often overlooked aspect of active system, but developing methods to systematically control the rotational dynamics of the active colloids has important implications for directed self-assembly.  Interestingly, hydrodynamic interactions between active colloids have been shown to have major effects on the their rotational dynamics, and the local field inducing the particle's propelling forces (pushers vs pullers) play a central role in MIPS \cite{zottl_hydrodynamics_2014,matas-navarro_hydrodynamic_2014,li_effects_2015,theers_modeling_2016,dunweg_lattice_2009,blake_spherical_1971,lighthill_squirming_1952,alarcon_spontaneous_2013}.

The notion that activity alone can generate giant number fluctuations and induce phase separation has important implications for self-assembly. In many colloidal self-assembly processes, the growth of a particular target structure requires the formation of a critical nucleus of a characteristic size. By introducing activity, it is then feasible to generate large density fluctuations that could be systematically exploited to extend the region in parameter space where self-assembly will successfully occur. In addition, self-propulsion will improve the kinetics of self-assembly as the local dynamics becomes inherently faster and, in more general, can make it easier for the system to escape local kinetic traps.  These issues will be discussed in greater detail in a later section. 
 
\subsection{Isotropically attractive ABPs}

\begin{figure}[t]
 \centering
    \includegraphics[width=0.7\textwidth]{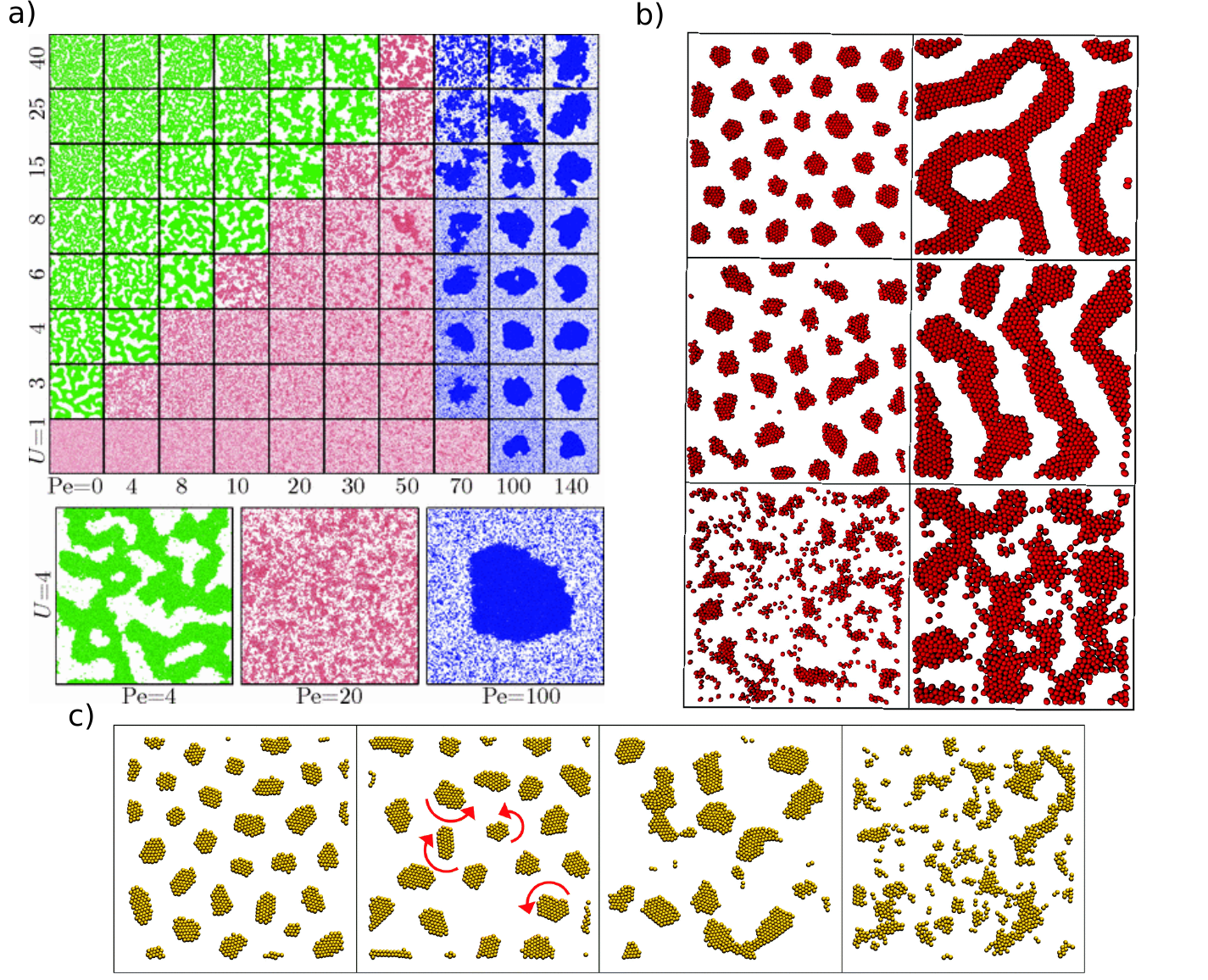}
  \caption{(a) Top panel: Phase behavior of isotropically attractive ABPs at an area fraction $\phi=0.4$ presented as a function of interparticle attraction strength $U$ and Peclet number $Pe=(v_p \sigma)/D$. The colors aid to distinguish the different phases: near-equilibrium gel states (green/upper-left), single-phase active fluids (red/center), and MIPS states (blue/right). (a) Bottom panel: Detailed snapshot at fixed interaction energy $U=4k_{\rm B}T$. At Pe=4 (left), the system  forms a kinetically arrested attractive gel. At $Pe=20$ (center), the attraction suppresses phase separation and produces a homogeneous fluid characterized by transient large density fluctuations. At $Pe=100$ (right) MIPS is observed.  (Panel reproduced with permission from American Physical Society from \cite{redner_reentrant_2013}.  (b) Snapshots of spherical ABPs interacting via a short-range attractive and a long-range resulsive potential at two different volume fractions $\phi = 0.24$ (left) and $\phi = 0.55$ (right). From top to bottom: Pe = 0, 10, 20. At Pe = 10, meso-phases of living crystals are observed at lower densities, and living stripes are seen at higher densities.  (c) Snapshots of self-propelled dumbbells interacting via a short-range attractive and a long-range resulsive potential at $\phi = 0.24$. From left to right we observe at Pe = 0 (static clusters), at Pe = 5 (rotating clusters), at Pe = 15 (living clusters) and at Pe = 40 (fluid phase). Panels \textit{b} and \textit{c} reproduced with permission from The Royal Society of Chemistry from \cite{tung_micro-phase_2016}}.
  \label{fig:fig3}
\end{figure}

The simplest extension over purely repulsive interactions is to consider spherical active colloids interacting via a short-ranged isotropic attraction.  Interestingly, and to some controversy \cite{ball_particle_2013}, the interplay between attraction and self-propulsion gives rise to new phenomenological behavior not observed in the passive parent system or an active system with purely repulsive interactions. In a recent article by \cite{redner_reentrant_2013}, the authors consider a two dimensional suspension of attractive active particles at an area fraction of $\phi=0.4$. The short ranged isotropic interaction is realized by allowing the ABPs to interact via a Lennard-Jones potential \cite{jones_determination_1924}. At this density, the phase diagram for this system exhibits reentrant behavior as a function of activity \cite{redner_reentrant_2013}. The phase diagram is given as a collection of simulation snapshots in Fig.\ref{fig:fig3}a. Self-propulsion can now either compete with the inter-particle attractions to suppress phase separation or act cooperatively to enhance it. At low levels of activity, the isotropic attraction between colloids dominates and the suspension phase separates in a similar manner to what expected in the parent equilibrium system.  In this regime, self-propulsion has little to no relevance.  At moderate levels of activity, the suspension melts into a single phase active fluid.  As the activity is increased even further, phase-separation driven by the self-trapping mechanism of MIPS will occur.  Here, energetic attractions act cooperatively with self-trapping to enable phase separation at lower propelling velocities than would be possible in the corresponding system with purely repulsive interactions.

Interestingly, experimental studies of active colloids \cite{palacci_living_2013,theurkauff_dynamic_2012,buttinoni_dynamical_2013,ginot_nonequilibrium_2015} at lower densities than the one discussed above \cite{redner_reentrant_2013}, observe the formation of dynamic clusters (living crystals) were each finite-sized cluster continually breaks, merges, dissolves, and re-forms.  This peculiar finding has also been observed in
two dimensional numerical simulations of self-propelled hard disks using kinetic Monte Carlo simulations 
\cite{levis_clustering_2014}, self-phoretic Active Brownian particles \cite{pohl_dynamic_2014},  and in a recent numerical study of a low density suspension of three dimensional attractive active colloids \cite{mognetti_living_2013,prymidis_state_2016}. These living aggregates appear to  form whenever attractive  and propelling forces are comparable in magnitude. A recent study of attractive squirmers (which account for explicit hydrodynamic interactions) has also observed system coarsening or clustering depending on the ratio between attraction and propulsion strength\cite{alarcon_morphology_2017}. 

Living clusters are a beautiful example of self-assembly into structures that can be active themselves. Unfortunately, it is not easy to control the size of these clusters using exclusively short range attractions, however, it is  possible to overcome this problem by adding to each particle a weak, but long-range repulsion. This creates a pair of competing forces at different distances, and it is known to drive  micro-phase separation into ordered structures of controllable size in passive suspension. A well-studied case is that of a short-range attraction (typically induced by depletion forces) coupled to a long-range repulsion (obtained, for instance, by giving the colloids a net charge). The result is the formation of a range of self-assembled structures whose morphologies depend on the balance between the attractive and repulsive interactions. At equilibrium and at relatively low  densities, this interplay can lead to the formation of isotropic clusters  arranged into hexagonal arrays  (or stripes at larger densities) whose size is to some degree tunable with the interaction parameters \cite{sciortino_equilibrium_2004,imperio_microphase_2006,fernandeztoledano_colloidal_2009,lu_gelation_2008}.

As far as we are aware, there are no experimental realizations of active systems of this variety. Nevertheless, two recent theoretical studies of two dimensional suspensions of ABP  interacting via a micro-phase separation-inducing long range potential have been carried out \cite{tung_micro-phase_2016,mani_effect_2015}. The potential used consisted, apart from the excluded volume, of  two generic parts: a short range attraction needed to induce clustering and a long range soft repulsion to guarantee micro-phase separation.  These studies have shown that for the appropriate choice of the inter-particle interactions and for a range of active forces, the living crystals of spherical ABP can be size-stabilized (See Fig. \ref{fig:fig3}b).  In the article by Tung et al. \cite{tung_micro-phase_2016}, the behavior of active dumbbells was also studied using the same inter-particle potential, and it was shown that it is possible to create mesoscopic crystalline structures that behave as micro-rotors (with definite, but tunable angular velocity). This occurs because of the particles' geometry  (See Fig.\ref{fig:fig3}c). The dumbbells tend to exhibit local nematic ordering upon aggregation causing the mesoscopic lattice sites to spin with a characteristic angular velocity. A similar behavior has been also observed in a two dimensional dilute suspension of self-propelled dumbbells interacting via a short-range attraction \cite{suma_motility-induced_2014}

The above  result allows us to make an interesting comparison between spherical and aspherical active colloids.  For spherical colloids interacting through an isotropic potential, the aggregates that develop are  intrinsically plastic, i.e. each particle within a structure is able to freely rotate, and this key feature is mainly responsible for the emerging active behavior of the resulting mesoscopic aggregates 
(e.g. living crystals). A rather different phenomenology is observed when considering particles endowed with directional interactions as was illustrated for the dumbbells in the previous example. These can develop either because of a non-spherical geometry of the particle or as a result of explicit anisotropic interaction potentials.  A significant amount of work has been done on the former case, with specific focus on isotropic self-propelled rods. The literature on the subject is extensive and focuses heavily on the behavior of swimming bacteria and active nematics (see for instance \cite{marchetti_hydrodynamics_2013} and references therein). In what follows, we review recent exciting results obtained on the self-assembly of spherical active Janus particles, which represent the simplest system with anisotropy in the inter-particle interactions.

\subsection{Dipolar Janus ABPs}

Dipolar Janus colloids consist of spherical particles coated with two oppositely charged hemispheres. The  directional interactions between these particles depend on the salt concentration in solution and allow for self-assembly of complex structures. In a recent article Kaiser et al. \cite{kaiser_active_2015} studied the clustering behavior of a small number of dipolar ABPs. The model considered by \cite{kaiser_active_2015} is similar to the ABP model introduced in the previous section with the addition of a point dipolar potential at the center of each colloid aligned with the self-propelling axis. In this work, the authors allowed a small number of dipolar ABPs  $(N=1-5)$ to self-assemble under equilibrium conditions and then introduced self-propulsion to understand its effect on the clusters' dynamics . This approach provides some intuition on how activity  induces structural changes in small clusters. 

A number of interesting features  having potentially important applications in colloidal self-assembly have been observed. For instance, dynamic structures  which are not stable at equilibrium develop when the particles are activated, and it is possible to both stabilize and destabilize particular configurations by the careful choice of the propelling velocity $v_p$. Interestingly, the stability of the equilibrium clusters is dependent on the ramp rate at which the self-propelling force is applied.  If self-propulsion is applied instantaneously, the clusters undergo a permanent fission process, where they either experience significant internal rearrangements or completely break. However, if self-propulsion is slowly ramped to its final value (adiabatic switching) no fission is observed and the equilibrium cluster configuration is retained.  This sensitivity to the ramp rate offers a unique handle for modulating self-assembly.  The authors illustrate a high level of control over the selection of a particular cluster configuration by tuning both the system parameters and by selecting an adequate choice of how self-propulsion is applied to the particle.  In a more recent paper the dynamical properties of these dipolar Janus ABP clusters were further characterized  \cite{guzman-lastra_fission_2016}.

Using a combination of numerical simulations and experiments, Yan et al. \cite{yan_reconfiguring_2016} have shown how multiple modes of collective behavior can be encoded into Janus dipolar ABPs.  In this work, each of the two hemispheres of the colloids carries its own electric charge and a colloid is propelled along the axis perpendicular to the particle's equator. The experimental realization of this system consists of silica spheres with one hemisphere coated with a metal and then other covered with a thin $SiO_2$ protective layer. The two hemispheres of the colloid polarize differently when an AC electric field is applied perpendicular to system. This mismatched frequency-dependent dielectric response is exploited to both control the motility of the colloids and modulate colloid-colloid interactions. 
\begin{figure}[t]
 \centering
    \includegraphics[width=0.5\textwidth]{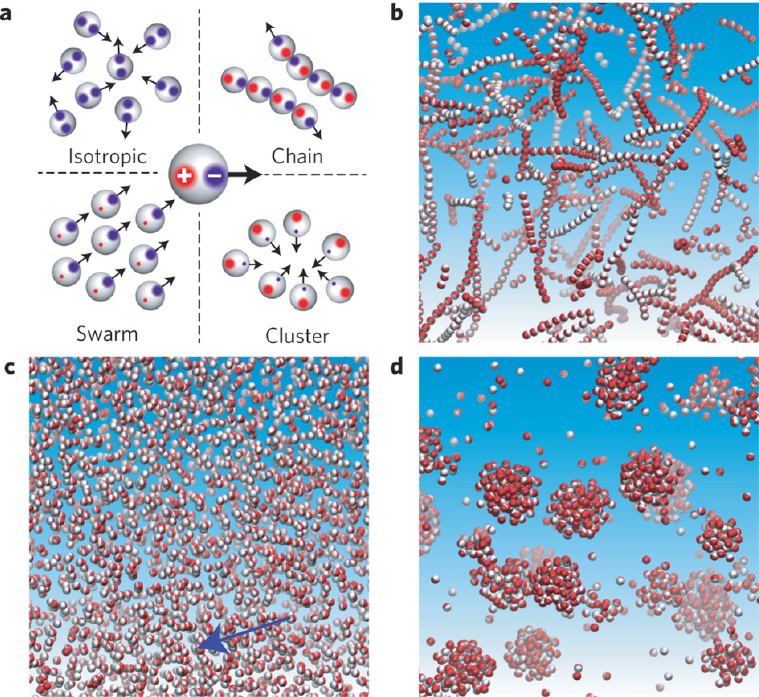}
  \caption{(a) Dipolar Janus ABPs where both the sign and magnitude of the charge on each hemisphere can be controlled to impose different collective behavior. The sign of the charge (positive and negative) are color coded (red and blue), their relative magnitude is indicated by the respective dot size, and the black arrows indicate the propelling directions. Panels (b) to (d) show typical snapshots for the different phases, including (b) active chains, (c) swarms and (d) clusters. The white hemispheres represent the leading sides and the red hemispheres the trailing sides. The blue arrow in panel (c) marks the global direction of the swarm. Image reproduced with permission from Nature \cite{yan_reconfiguring_2016}.}
  \label{fig:fig4}
\end{figure}
As shown in Fig.\ref{fig:fig4}, it is possible to self-assemble several distinct active structures by  modulating  magnitude and sign of the electric charge on each hemisphere of the colloid.  When the charge on the two hemispheres has the same sign and magnitude, the system exhibits behavior similar to that of purely repulsive ABPs, and since the density considered is below the onset for the MIPS transition, the system remains in a fluid state (Fig.\ref{fig:fig4}a). More interestingly, when there is a charge imbalance between the two hemisphere, the system develops various forms of non-trivial self-organization. Particles with equal and opposite charges on the two hemispheres spontaneously assemble into  active chains (Fig.\ref{fig:fig4}b). When a large charge imbalance between the two hemispheres is enforced, the colloids induce large torques on one another when  coming in close proximity, and two distinct cases emerge. Case (1): when the propelling force is set to point towards the hemisphere with greater charge, the torques tend to aligns two colloidal swimmers in the same direction. This angular alignment produces large scale coherent swarms with phenomenology reminiscent of the Vicsek model (Fig.\ref{fig:fig4}c) \cite{vicsek_novel_1995}. Case (2):  when the propelling force is set to point away from the hemisphere with greater charge, the induced torques tend to rotate the colloids so that the more charged hemispheres are pushed as far apart as possible from each other.  In this case the  colloids tend to align with their propelling axis facing one another resulting in the formation of jammed clusters with high local density (Fig.\ref{fig:fig4}d).  These studies provide an excellent example of how angular interactions can be tuned to induce  self-assembly of complex structures, and indicate how the interplay between the  torques and the self-propelling forces can facilitate the formation of active aggregates.

\subsection{Amphiphilic Janus ABPs}

\begin{figure}[h!]
 \centering
    \includegraphics[width=0.8\textwidth]{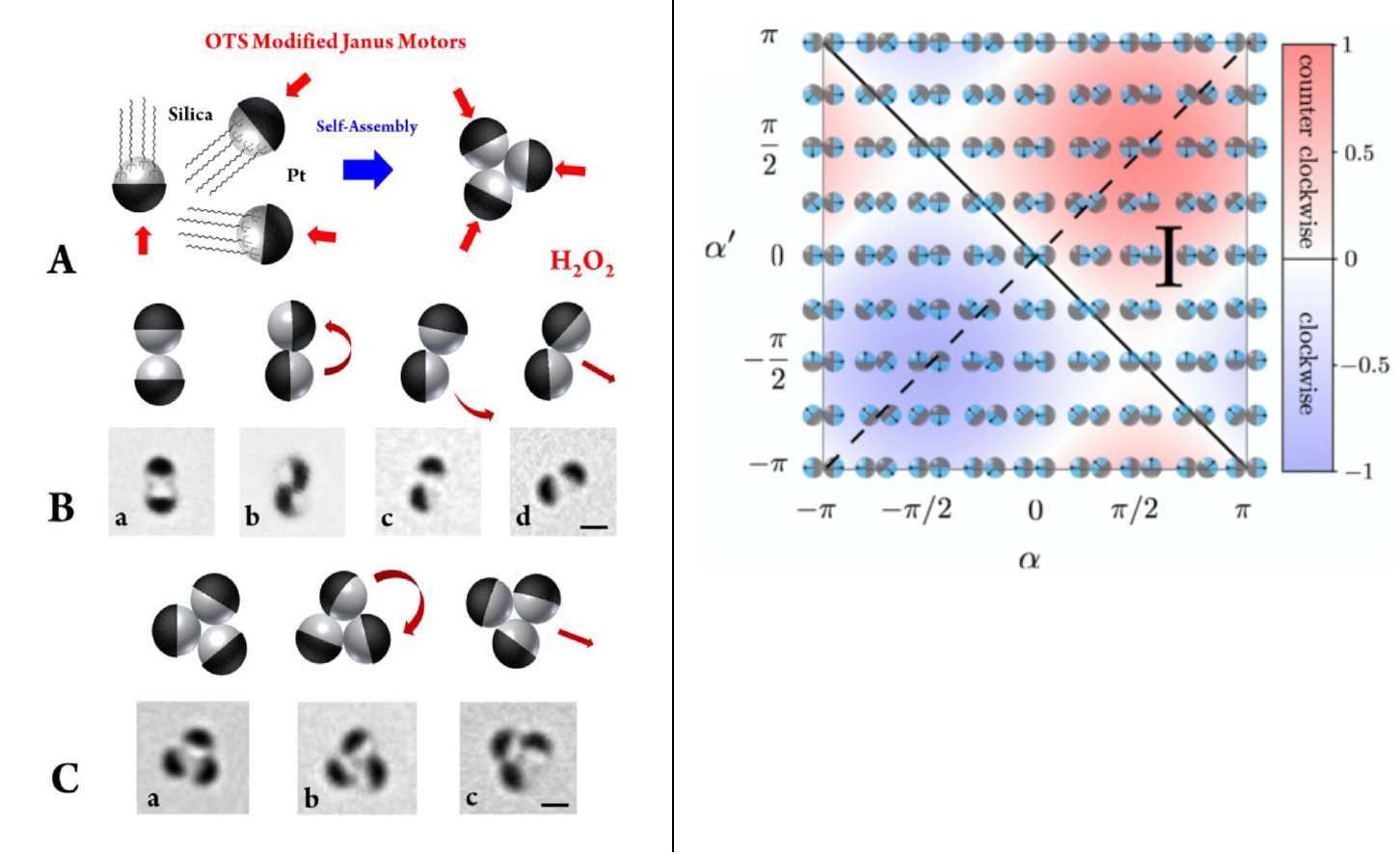}
  \caption{Left panel: (A) Sketch of the assembly of OTS modified amphiphilic Janus ABPs from \cite{gao_organized_2013}, with the hydrophobic OTS sides (in gray) facing one another. (B) Typical configurations (a$-$d) of a pair of amphiphilic Janus ABPs (Scale bar, 1 $\mu$ m). (C) Typical assemblies of amphiphilic Janus ABPs triplet assemblies: (a$ - $c) movement of triplet assemblies with different orientations (Scale bar, 1 $\mu$ m). Reproduced with permission from the American Chemical Society from \cite{gao_organized_2013}. Right panel: Plot representing the angular speed of possible dimer conformations. The intensity of the shaded regions show counterclockwise (red) and clockwise (blue) rotations of the dimers. Reproduced with permission from American Physical Society from \cite{johnsondynamic2017} }
  \label{fig:fig5}
\end{figure}

We now shift our focus to the self-assembly behavior of amphiphilic Janus ABPs. This type of Janus particle consists of two distinct hemispheres, one hydrophobic and the other hydrophilic. An experimental realization of an amphiphilic Janus ABPs consists of silica microspheres passivated with a hydrophobic ligand (octadecyltrichlorosilane, OTS), whose hemisphere is then covered by a Pt cap. These amphiphilic Janus ABPs experience a short ranged attraction between the hydrophobic domains, while the interaction between the Pt capped hemispheres is purely repulsive. In essence, the Pt cap restricts the inter-particle attraction to the hydrophobic hemisphere which gives rise to well defined assemblies (See Figure \ref{fig:fig5}).  By tuning the size of the hydrophobic domain and that of the Pt cap, it is possible to control the angular range of the interaction, and thus the number of colloids that can simultaneously stick to each other, and the degree of self-propulsion of the the colloid.

Gao et al. \cite{gao_organized_2013}, investigated experimentally the formation and dynamics of small clusters of amphiphilic Janus ABPs. For a pair of colloids. They find that the motion is governed by the relative orientation of the two self-propelling force. Figure \ref{fig:fig5} enumerates the possible orientations for a pair of amphiphilic Janus ABPs.  When the two propelling vectors are aligned with the interparicle axis, so that the particles directly push against each other,  the pair  becomes essentially passive.  The two active forces  cancel each other and the motion of the  pair is purely Brownian. However, when the propelling vectors are both perpendicular to the interparticle axis, the pair behave as a colloidal rotor with a characteristic angular velocity determined by the magnitude of the self-propelling force.  The self-propelling force in this case only contributes to the rotation of the pair.  The translational dynamics of the pair are effectively Brownian when this configurations is adopted. When the location of the Janus balance is altered so that the size of the hydrophobic patch becomes smaller, the rotational speed of the pair decreases and it revolves with a larger radius. While it is straightforward to estimate the degree of translational and rotational motion for a particular configuration of a pair of Janus amphiphilic ABPs (as shown  in some detail in ref. \cite{johnsondynamic2017}) the situations becomes much more complex when dealing with aggregates of a large number of particles. 

One of the biggest challenges in active self-assembly is designing a building block  where the interparticle interactions and the self-propulsion of the particles work in conjunction to bias the formation of a particular target structure.  The next section in this review is dedicated to this topic.

  \section{\label{sec:reverse} Can Activity Improve Self-Assembly?}

So far we have discussed some of the new exciting behavior observed in suspensions of active colloids with a specific focus on the new properties, morphological or dynamical, of the aggregates formed under the influence of propelling forces. In this last section, we shift the focus, and consider whether activity can be exploited to speed up the complex self-assembly dynamics of colloids designed to (passively) form desired target structures. For this purpose,  we will consider activity as a new, tunable dimension in the colloidal parameter space.  This new variable, self-propulsion,  as already illustrated in the previous sections, can profoundly affect the dynamical behavior of a system. However, when implemented in a synergistic manner, it can be exploited to bias the formation of a particular desired microstructure against any malformed or competing microstructures. 

As discussed in the introduction, it is quite clear that even when the interactions between building blocks are chosen so that a particular ordered structure is stable, the self-assembly of that structure may not be possible due to the presence of competing metastable states.  Successful self-assembly  usually requires a delicate balance between entropic and dispersion forces, and the formation of defect-free structures demands that the rate of association of the building blocks to the forming assembly to be comparable to the rate of dissociation from it, leading to rather slow time scales for the completion of the self-assembly process. Obviously, because of its enhanced translational motion,  self-propelled particles can  explore a given volume and find a large slow target (the assembly) much faster than their passive counterparts. However, the same propelling forces, when sufficiently large can alter the morphology, the structural properties, and even break down the target structure by exerting unbalanced inner forces and torques. In other words, if on the one hand self-propulsion will speed up the kinetics of particle association, on the other hand it can affect the stability of the desired target structure. As such, one should not expect that adding self-propulsion to any colloidal particle will necessarily be advantageous, unless the self-propelling forces themselves will organize within the target assembly in such a way that its stability is not compromised. This can be achieved by judiciously pairing the propelling forces and the inter-particle interactions.

\begin{figure}[h!]
 \centering
    \includegraphics[width=0.8\textwidth]{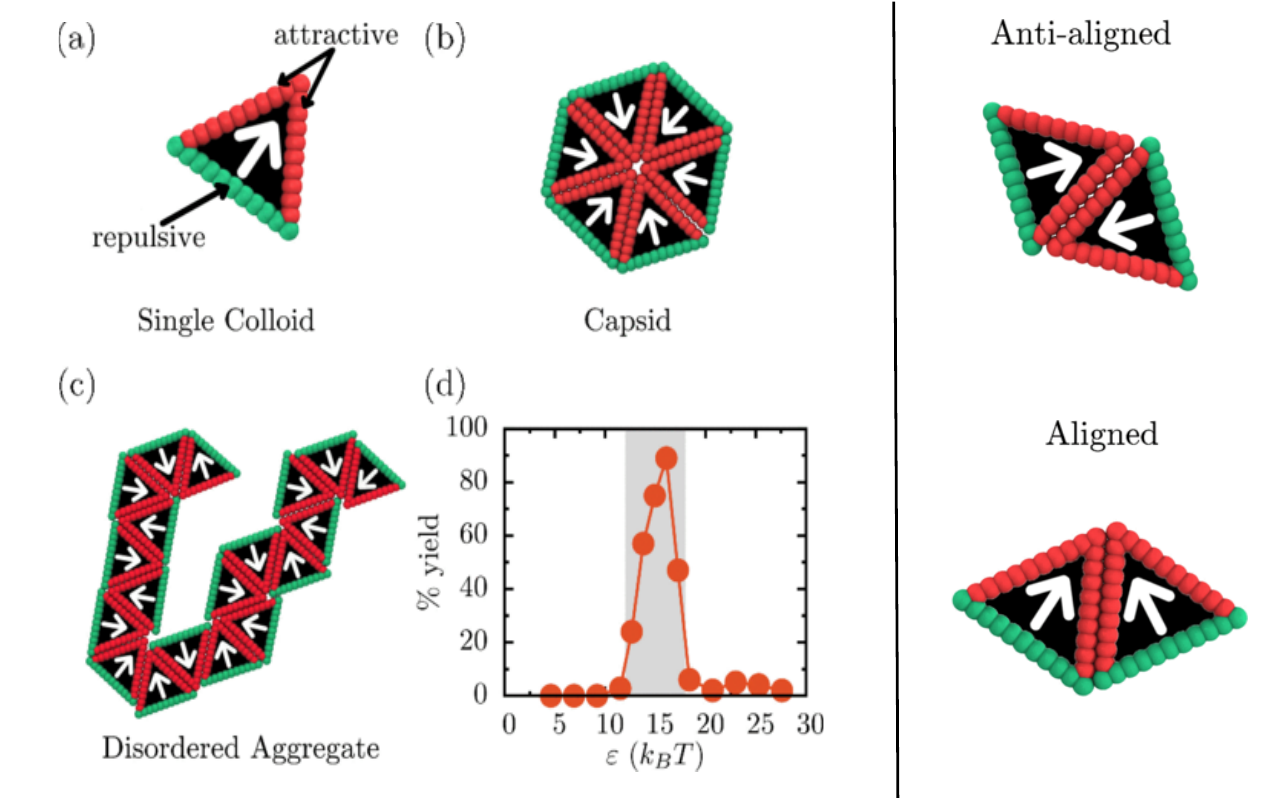}
  \caption{Left panel: (a) Sketch of a triangular  building block with two attractive sides. The white arrow indicates the direction of self-propulsion. (b) The target structure to be self-assembled is a hexagonal capsid composed of six colloidal building blocks. (c) Typical disordered aggregate formed when the attraction between the triangles is too strong. (d) Percent yield of target structure as a function of the strength of the attractive interaction for passive colloids. Right panel: Anti-aligned (top) and aligned (bottom) configurations of a pair of triangular particles.}
  \label{fig:fig6}
\end{figure}

As a simple illustration, we review the self-assembly of active triangular colloidal particles discussed in ref. \cite{mallory_activity-assisted_2016}.  Although there is nothing particularly special about the shape of this building block, the pathway associated to its self-assembly presents all the generic features observed in a typical self-assembly process:(1) there is a clearly defined target structure. Equilateral triangles with one side hard and the other two attractive are designed to passively self-assemble into hexagonal structures of exactly six triangles (capsids) [see Fig.\ref{fig:fig6}a]. (2) There exist an ensemble of structures with a larger degree of orientational entropy [see Fig.\ref{fig:fig6}(c)] that can directly compete with the formation of the target structure as soon as the inter-particle interaction becomes sufficiently large. This high level of competition generates a scenario where the success of self-assembly of  the hexagonal structures is highly dependent on the strength of attraction between colloids and is expected to only occur for a narrow energy range. (3) Starting from a low density suspension, the typical times required for self-assembly becomes exceedingly slow. 

In Fig.\ref{fig:fig6}d, the percent yield of the target structure as a function of the strength of the attractive interaction is given. As expected, successful self-assembly in the passive system is limited to a narrow range in the binding 
energy.  In this window, the maximum yield observed is $\sim 90 \%$.  For smaller values of the binding energy, the colloids do no aggregate. For larger values, the formation of large disordered aggregates is favored [see Fig.\ref{fig:fig6}c]. How can one endow these particles with self-propulsion to improve the overall rate and yield of self-assembly?

The general strategy in implementing this approach is to design the colloidal building blocks such that the particle interactions and the direction of self-propulsion work in conjunction to form and stabilize the desired structure. In this simple example system, the solution is fairly straightforward.  By choosing the direction of self-propulsion to be perpendicular to the repulsive face of the colloid as illustrated in Fig.\ref{fig:fig6}a, one can  improve the stability of the final hexagonal aggregate. Furthermore, this choice biases the formation the target strucure early on in the self-assembly process. Indeed, as illustrated in  Fig.\ref{fig:fig6}(left panel), there are only two ways for a pair of colloids to bind to each other. In the first configuration, the propelling axes are anti-aligned, whereas in the second configuration the self-propelling axes are partially aligned. In the former case, activity tends to destabilize and separate the pair while in the latter case the bond between the two particles is strengthened by the contributions of the active force.  Biasing the system away from binding configurations that lead to the formation of misaligned structures leads to a significant increase in the likelihood of a robust self-assembly process. Thus the selection of the final structure begins already at an early stage, as  partially aligned configurations are favored and are compatible with the target geometry.  

A critical requirement of the target microstructure is that the vectorial sum of the self-propelling forces in its interior is equal to zero. This creates a focal point in the center of the compact aggregate where each colloid can exert a force  that strengthens their mutual attractive interactions through the cooperative arrangement of the propelling forces. These
emerging interactions stabilize the hexagonal structure even in the absence of explicit attractive forces and behave essentially as {\it active bonds} of tunable strength. Any compact aggregate for which this vectorial condition is not satisfied will experience large active torques and shear forces that can break them apart. One can think of self-propulsion in these systems as a very selective filter that only allows for the stabilization of certain structures. We believe that this strategy should work for all compact target structures satisfying the vectorial condition discussed above.


We should stress that this behavior also suggests ways of using activity as a tool to separate at once both time and energy scales in multi-stage self-assembly problems \cite{patteson_active_2016}. In fact, given that  the aggregation rate and the strength of the active bonds are completely controlled by the particles’ self-propulsion, it is possible to use activity to quickly self-assemble compact structures with zero net propulsion, such as the hexagonal capsids described above, that would function as new passive mesoscopic building-blocks. These could then be made to slowly  self-assemble even further exploiting standard dispersion or weak solvent-induced interactions. The ability to efficiently separate energy and time scale within the same sample can be greatly beneficial for the fabrication of evermore-complex microscopic architectures.

Although this strategy appears to be rather effective for compact but finite-sized structures, there is evidence \cite{mallory_improving_nodate} that it can also be applied to other systems. For instance,  low levels of activity can be used to improve the self-assembly of macroscopic crystals and open lattice structures. It has been shown \cite{chen_directed_2011} that passive triblock Janus particles, which have two attractive patches located at the opposite poles of each particle, will spontaneously form macroscopic kagome lattices for sufficiently large patch areas and attraction strengths. Our preliminary results \cite{mallory_improving_nodate} indicate that the formation of the kagome lattice structure greatly benefits from the introduction of  a small amount of self-propulsion along the pole-to-pole axis of these particles. The reason behind this is that the active forces in this case act to destabilize the competing elongated structures that tend to form early on in the self-assembly process. Clearly more work in this direction  needs to be done to understand the intricacies of how activity may affect the stability or the dynamic pathway of the self-assembly process, nevertheless this is a very promising direction in the field.

\section{\label{sec:conclusion}Conclusions}
In this review, we recapitulated some of the most exciting phenomenological behavior observed in suspensions of active colloids. We also discussed some of the strategies that can be deployed to exploit activity to more efficiently self-assemble target structures. To conclude, we would like to remark that apart from a few exceptions (see for instance \cite{stenhammar_light-induced_2016,bianchi_active_2016} and references therein), most of the work in this area has been carried out under the assumption that the active forces are constant with respect to time.  However, the newest generation of active colloids, can  be light activated \cite{buttinoni_active_2012,palacci_living_2013,palacci_light-activated_2014} (i.e. their self-propulsion can be turned on and off by exposure to an external light source). This is achieved, for instance, by inserting a piece of hematite on one side of the colloids and shining light of a specific frequency on the suspension $-$ it is only in these illuminated regions that the hematite catalyzes the hydrogen peroxide \cite{palacci_living_2013}.  This is an incredibly useful feature for self-assembly as it makes it feasible to create regions and sculpt patterns in the sample where particles become active (under light exposure) and regions where particles behave as purely Brownian objects. Another interesting feature of light-induced activity is that it allows the strength of the active forces to be easily modulated over time by either tuning the intensity of the light source or by switching it on and off at different frequencies. This simple feature opens the way to experimental studies of stochastic Brownian ratchets, and adds yet another handle that could be exploited to create new functional materials, allowing one to directly intervene and dynamically affect the pathway of structure formation.


\bibliographystyle{unsrt}


\end{document}